\documentclass[aps,pra,twocolumn,superscriptaddress,showpacs]{revtex4}

\usepackage{amsmath}
\usepackage{graphicx}
\usepackage{bbm}

\newcommand{\ket}[1]{\ensuremath{|#1\rangle}}

\newcommand{\ketbra}[1]{\ensuremath{| #1 \rangle \langle #1 |}}
\newcommand{\braopket}[3]{\ensuremath{\langle #1|#2|#3\rangle}}
\newcommand{\braket}[2]{\ensuremath{\langle #1|#2\rangle}}

\newcommand{\expect}[1]{\ensuremath{\langle #1 \rangle}}
\DeclareMathOperator{\Trace}{tr}
\newcommand{\tr}[1]{\ensuremath{\Trace\left(#1\right)}}
\newcommand{\sqnorm}[1]{\ensuremath{\left| #1 \right|^2}}

\newcommand\vect[2]{\left(\begin{array}{#1}#2\end{array}\right)}

\newcommand\Unit{\ensuremath{\mathbbm{1}}}
\newcommand\Ham{\mathcal{H}}
\renewcommand\P{\mathcal{P}}
\renewcommand\S{\mathbf{S}}

\newcommand\refeq[1]{Eq.~(\ref{eq:#1})}
\newcommand\refq[1]{(\ref{eq:#1})}

\newcommand\refapp[1]{Appendix~\ref{sec:#1}}
\newcommand\reffig[1]{Fig.~\ref{fig:#1}}
\newcommand\reftable[1]{Table~\ref{table:#1}}

\begin{document}

\title{Quantum pattern recognition with liquid-state nuclear magnetic resonance}

\author{Rodion Neigovzen}
\affiliation{Siemens AG, Corporate Technology, Otto-Hahn-Ring 6, D-80200 Munich, Germany}
\affiliation{Department of Physics, Technische Universit{\"a}t M{\"u}nchen, James Franck Str., D-85748 Garching, Germany}
\author{Jorge L. Neves}
\affiliation{Department of Chemistry, Technische Universit{\"a}t M{\"u}nchen, Lichtenbergstrasse 4, D-85747 Garching, Germany}
\author{Rudolf Sollacher}
\affiliation{Siemens AG, Corporate Technology, Otto-Hahn-Ring 6, D-80200 Munich, Germany}
\author{Steffen J. Glaser}
\affiliation{Department of Chemistry, Technische Universit{\"a}t M{\"u}nchen, Lichtenbergstrasse 4, D-85747 Garching, Germany}

\date{\today}

\begin{abstract}
A novel quantum pattern recognition scheme is presented, which combines the idea of a classic Hopfield neural network with adiabatic quantum computation. Both the input and the memorized patterns are represented by means of the problem Hamiltonian. In contrast to classic neural networks, the algorithm can return a quantum superposition of multiple recognized patterns. A proof of principle for the algorithm for two qubits is provided using a liquid state NMR quantum computer.
\end{abstract}

\pacs{03.67.Lx, 03.67.Ac, 07.05.Mh, 82.56.--b}
\keywords{quantum computation, adiabatic quantum computation, artificial intelligence, Hopfield neural nets, pattern recognition, quantum computational intelligence, nuclear magnetic resonance}

\maketitle

\section{Introduction}

The framework of Natural Computing tries to model architectures found in Nature and to apply them to a diversity of computational tasks. Its biological domain is represented among others by neural networks and evolutionary computation. In physics, novel algorithmic schemes have been investigated in the context of quantum computation. Based on these ideas, we investigate how pattern recognition can be implemented using principles of quantum computing.

Artificial neural networks involve interacting units called neurons designed according to neural structures of a brain which cooperate in order to process information \cite{Haykin1998}. Similar to the brain, artificial neural networks are capable of performing cognitive tasks such as pattern recognition and associative memory. Pattern recognition processes input data usually on the basis of \textit{a priori} knowledge in the form of a set of memorized patterns.
An input pattern is then classified to whichever one of the memory patterns it most closely resembles. Typical applications for pattern recognition include automatic recognition of objects and patterns in digital image analysis as well as voice and speech recognition. An associative memory is a system that completes \textit{partially} known input pattern based on the stored content. Typical applications for associative memory include content-addressable memory as a special type of computer memory and database engines.

A classic theoretical model to perform the described tasks is a Hopfield network \cite{Hopfield1982}, a recurrent neural network with symmetric connections between individual neurons. Full connectivity is provided with every neuron $i$ being able to interact with any other neuron $j\neq i$ by means of a response function $r_i(t)=\sum_{j\neq i} w_{ij} S_j(t)$ aggregated as a weighted sum over the current states of all other bipolar neurons $S_j(t)=\pm 1$ which correspond to biological states of not firing and firing electrical signals to its neighbor neurons. An activation function $f_i(x)$ of the neuron $i$ then evaluates $r_i$ in order to define if the neuron remains in its currents state or a state flip is applied, e.g., $S_i(t+\Delta t) = f_i(r_i(t))$, where $f_i(x)=sgn(x)$ is the sign function.

For a set of $p$ patterns $\P=\{\xi^1,\ldots,\xi^p\} = \lbrace \xi^{\mu}\rbrace$ with $\mu=1,\ldots,p$ and bipolar $\xi_i^\mu = \pm 1$ different definitions for the synaptic connection strength $w_{ij}$ between neurons $i$ and $j$ are possible, e.g. by the (discrete) Hebbian matrix
\begin{equation}
\label{eq:hebbian_matrix}
w_{ij}=\frac{1}{N}\left[\sum _{\mu=1}^p \xi_i^{\mu}\xi_j^{\mu}-p\delta_{ij}\right],
\end{equation}
which stores the memory information in a distributed manner.
The input binary vector $\xi^{inp}$ is imposed on the Hopfield network as its initial state. The dynamics of the system is designed such as to minimize a cost function
\begin{equation}
\label{eq:cost_function}
E(\S(t),w)=-\frac{1}{2}\sum_{ij}w_{ij}S_{i}(t)S_{j}(t)
\end{equation}
by converting the state of the network to a stable configuration which represents the recognized pattern. Thus, the physical background of the Hopfield model offers an approach for quantization of the neural network.

Quantum computers utilize special characteristics of quantum systems such as superposition and entanglement \cite{Mermin2007}. The corresponding algorithms provide computational effectiveness in comparison with classic routines for problems such as search \cite{Grover1997a} and factorization \cite{Shor1997}. Different hardware designs have been proposed for building a quantum computer, including liquid state NMR \cite{Cory1997, Gershenfeld1997, Vandersypen2002, Marx2000} and ion traps \cite{Cirac1995}. The extensive technological development of NMR spectroscopy in the last five decades made it possible for NMR quantum computing to become a suitable test ground for novel quantum algorithms.

A systematization of quantum computing using adiabatic evolution leads to the concept of Adiabatic Quantum Computation (AQC) \cite{Farhi2001} that has been proven to be equivalent to the standard network model of quantum computation \cite{Aharonov2004}. Compared to the abstract computational language of the quantum networks model, AQC often provides a more direct translation to experimental quantum computing.

The protocol for AQC is represented by a controlled Hamiltonian path $\Ham(s)$ with $s=\frac{t}{T} \in [0,1]$ and running time $T$. The computational problem is encoded in the final Hamiltonian $\Ham(1)=\Ham_p$ and the solution of the computational problem corresponds to finding the ground state of $\Ham_p$. On the other side, the initial state of the quantum system $\ket{\psi(0)}$ is chosen to be an easily preparable ground state of $\Ham(0)=\Ham_i$. If the Hamiltonian is driven from $\Ham_i$ to $\Ham_p$ slowly enough, i.e., if the total calculation time $T$ is chosen to be long enough, then in the adiabatic limit the final state of the quantum system comes arbitrarily close to the problem ground state. AQC can also be formulated in the framework of Quantum Annealing \cite{Santoro2006} utilizing a Hamiltonian trajectory
\begin{equation}
\label{eq:qanneal_trajectory}
\Ham(s) = \Lambda(s) \Ham_i + \Ham_p
\end{equation}
with $\Lambda(0)$ large enough to make $\Ham_i$ the dominating term and $\Lambda(1)=0$ which reduces the Hamiltonian to the constantly present contribution by $\Ham_p$. The transverse Hamiltonian $\Ham_i$ represents the driver for the adiabatic evolution whereas the initial state $\ket{\psi(0)}$ of the annealing process is chosen so it corresponds approximately to an energy eigenstate of the initial Hamiltonian $\Ham(0) = \Lambda_{max}\Ham_i + \Ham_p$.

In classic terms, AQC represents minimization of an energy cost function. This offers a link from (usually irreversible) classic neural dynamics to quantum dynamics governed by unitary, reversible evolution. However, the classic Hopfield network approach performs pattern recognition by \textit{local} optimization. In the following, we utilize AQC which performs \textit{global} optimization in form of ground state approximation.

\section{Theory}

We consider a quantum neural network with $N$ neurons consisting of bipolar neural states $-1$ and $+1$ to be represented by a quantum system with $N$ qubits in states $\ket{0}$ and $\ket{1}$. Whereas a classic artificial neuron can assume only a single ``fire'' or ``not fire'' state at once, quantum neurons allow superpositions of these two states in the form of $\alpha\ket{0}+\beta\ket{1}$ with $|\alpha|^2+|\beta|^2=1$.

\begin{figure}
\includegraphics[width=8.5cm]{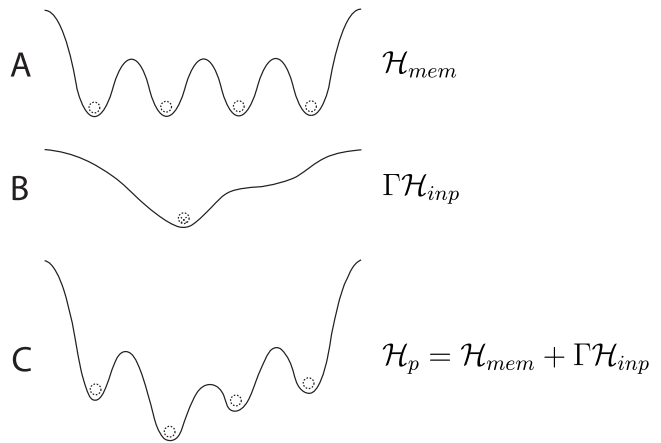}
\caption{Illustrative representation of the energy landscapes for the memory Hamiltonian $\Ham_{mem}$ (A), the weighted input Hamiltonian $\Gamma \Ham_{inp}$ (B) and the total problem Hamiltonian $\Ham_p$ (C). Dashed circles represent memory patterns stored as energy minima of $\Ham_{mem}$; crossed dashed circle represents the input pattern as energy minimum of $\Ham_{inp}$.}
\label{fig:problem_hamiltonian}
\end{figure}

In order to apply AQC to the computational task of (associative) pattern recognition, we encode the complete problem in the Hamiltonian
\begin{equation}
\label{eq:problem_hamiltonian}
\Ham_p = \Ham_{mem} + \Gamma \Ham_{inp}
\end{equation}
with $\Ham_{mem}$ representing the knowledge about stored patterns, $\Ham_{inp}$ representing the computational input and an appropriate weight factor $\Gamma>0$ (see \reffig{problem_hamiltonian}).

\begin{figure}
\includegraphics[width=8.5cm]{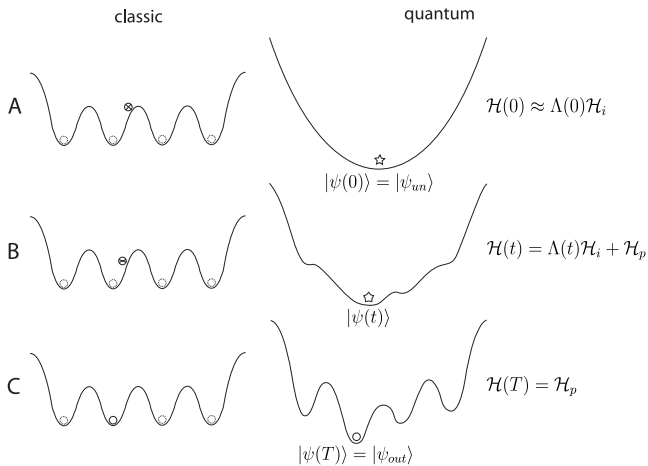}
\caption{Illustrative representation of classic and quantum dynamics within the energy landscape. Classic dynamics drives the initial state (crossed circle) which represents the input pattern toward the final state (solid circle) which represents a nearby memory pattern. Quantum dynamics transforms the energy landscape from the initial Hamiltonian close to $\Ham_i$ (A) via intermediate Hamiltonian $\Ham(t)$ (B) towards the final Hamiltonian equivalent to the problem Hamiltonian $\Ham_p$ (C). The current state of the system remains the ground state of the instantaneous Hamiltonian with the ``empty memory'' state $\ket{\psi_{un}}$ \refq{empty_memory_state} for $t=0$ (represented by the star in A) and the final state of the system corresponding to the outcome memory pattern for $t=T$ (represented by the solid circle in C).}
\label{fig:dynamics}
\end{figure}

Whereas the classic Hopfield network utilizes dynamics defined by the memory patterns alone, the dynamics according to \refeq{problem_hamiltonian} depends on both memory and input information (see \reffig{dynamics}).

Furthermore, the encoding of the final adiabatic Hamiltonian as a sum of two independent terms offers an advantage with respect to the practical implementation of the process.
Using appropriate approximations (such as Suzuki-Trotter), the computational evolution of the system can be approximated by separate periods of unitary evolution each according to the individual terms building the total instantaneous Hamiltonian. This implies that the memory precompilation, i.e., the translation of dynamics according to $\Ham_{mem}$ into elementary physical interactions such as pulse sequences in NMR quantum computing systems has to be executed only \textit{once} for an individual memory set. The generated low-level commands can then be reused for variable input information. The dynamics according to the input Hamiltonian $\Ham_{inp}$ can be obtained and imposed independently.

In direct analogy to energy function \refq{cost_function}, the memory Hamiltonian can be defined by the coupling strengths between the qubits
\begin{equation}
\label{eq:memory_hamilt_coupl}
\Ham_{mem} = -\frac{1}{2}\sum_{i\neq j} w_{ij} \sigma^z_i \sigma^z_j,
\end{equation}
where $\sigma^z_i$ is the Pauli $z$ matrix on qubit $i$ and $w_{ij}$ is the Hebbian matrix \refq{hebbian_matrix}.
A more general choice of the memory Hamiltonian is discussed in \refapp{projector_memory}.

For the retrieval Hamiltonian $\Ham_{inp}$ we consider a single input pattern $\xi^{inp}$ of length~$N$. For associative memory applications we can extend a noncomplete input vector $\hat{\xi}^{inp}$ of length~$n<N$ by setting the values for $N-n$ unknown states to zero. In contrast to the classic neural network, we impose the input pattern $\xi^{inp}$ on the dynamics of the system by an additional Hamiltonian term $\Ham_{inp}$ which can be defined as
\begin{equation}
\label{eq:retrieval_hamiltonian}
\Ham_{inp} = \sum_{i} \xi^{inp}_i \sigma^z_i.
\end{equation}
For a given state $\ket{\xi}$ representing a pattern $\xi$, the energy corresponding to $\Ham_{inp}$ is given by $E^{inp}(\xi) = - n + 2\hat{h}$ and depends on the Hamming distance $\hat{h}=0,\ldots,n$ between $\hat{\xi}^{inp}$ and $\hat{\xi}=[ \xi_1,\ldots,\xi_n ]$, i.e., the number of positions for which the corresponding entries are different. The external field defined by $\Ham_{inp}$ thus creates a scalar metric proportional to the Hamming distance between $\xi^{inp}$ and memory patterns.
This permits a quantitative comparison between input and memory patterns by shifting the energy levels of the memory Hamiltonian $\Ham_{mem}$ so the more similar memory patterns have lower energy compared to the alternative patterns. It can be shown that for the case of a single stored pattern ($p=1$), the upper bound for the weight factor in \refeq{problem_hamiltonian} is given by $\Gamma < 1-\frac{n}{2N}$ which ensures that the correct memory pattern is the ground state of the system (cf. \refapp{input_weight}).

Finally, we define the initial conditions of the AQC protocol. In comparison with the classic Hopfield network, the initial state of quantum pattern recognition systems does not correspond to the input pattern but is chosen as
\begin{equation}
\label{eq:empty_memory_state}
\ket{\psi(0)}\equiv\ket{\psi_{un}}=\frac{1}{2^\frac{N}{2}}\sum_{k=0}^{2^N-1}\ket{k}
\end{equation}
representing an ``empty memory'' with uniformly distributed probability for all possible state configurations.
As corresponding initial Hamiltonian, we choose an easily constructible sum over single qubit operators $\Ham_i = \frac{1}{2}\sum_i (\Unit - \sigma^x_i)$ \cite{Farhi2000a} which can be implemented experimentally in a straightforward way.
Since the problem Hamiltonian $\Ham_p$ \refq{problem_hamiltonian} is diagonal, the transverse form of $\Ham_i$ ensures the transformation of the initial state $\ket{\psi_{un}}$ \refq{empty_memory_state} toward the outcome memory pattern.

\section{Experiments}

For the simulations and experiments, we considered a heteronuclear two spin system corresponding to the $^{1}$H and $^{13}$C nuclear spins $\frac{1}{2}$ of $^{13}$C labeled sodium formate dissolved in D$_2$O. The experiments were performed at a temperature of 27$^\circ$C using a Bruker AC 200 spectrometer operating at a Larmor frequency of 200~MHz for $^{1}$H and 50~MHz for $^{13}$C. The heteronuclear $^{1}$H-$^{13}$C coupling constant is $J=195$~Hz and the relaxation times are $T_1(^1\textnormal{H})=1.6$~s, $T_1(^{13}\textnormal{C})=2.7$~s, $T_2(^1\textnormal{H})=130$~ms, and $T_2(^{13}\textnormal{C})=60$~ms. In this section, we use normalized Pauli operators as conventional in NMR \cite{Ernst1990}.

In order to provide an experimental demonstration for the concept of quantum pattern recognition, we implement a basic two-neurons Hopfield network using a liquid state NMR quantum computer.
The single connection $w=w_{12}=w_{21}$ between two qubits corresponds for $w=-1$ to patterns $[-1,+1]$ and $[+1,-1]$ with different bit values whereas $w=+1$ corresponds to patterns $[-1,-1]$ and $[+1,+1]$ with identical bit values. Following \refeq{problem_hamiltonian}, the problem Hamiltonian can be written as
\begin{equation}
\label{eq:twoqubit_hamiltonian}
\Ham_p = -w \sigma^z_1 \sigma^z_2 + \Gamma \left(\xi^{inp}_1 \sigma^z_1 + \xi^{inp}_2 \sigma^z_2\right).
\end{equation}
The NMR free evolution Hamiltonian
\begin{equation}
\label{eq:nmr_hamiltonian}
\Ham'_{p} = 2 \pi J \sigma^z_H \sigma^z_C + 2 \pi \nu_H \sigma^z_H + 2 \pi \nu_C \sigma^z_C
\end{equation}
for the considered molecule shows a constant positive coupling.
The variable memory weight~$w$ according to \refeq{twoqubit_hamiltonian} can be implemented by changing the sign of the effective coupling Hamiltonian using additional pulses \cite{Haeberlen1976, Jones1999a}.
In the case of two qubits considered here an alternative approach which avoids the need for additional pulses is the following. We can map the problem by defining the offset frequencies $\nu_{H} = - w \Gamma J \xi_1^{inp}$ and $\nu_{C} = - w \Gamma J \xi_2^{inp}$ and redefining the initial Hamiltonian as problem dependent $\Ham'_i = 2 \pi w ( \sigma^x_H + \sigma^x_C )$.
The initial quantum state $\ket{\psi_{un}}$ of \refeq{empty_memory_state} corresponds to the ground state of $\Ham_i$ for $w=-1$ and to its highest excited energy eigenstate for $w=+1$. Our choice of $\nu_{H}$ and $\nu_{C}$ consequently encodes the desired $\ket{\psi_{res}(w,\xi^{inp})}$ as the ground state, respectively, the corresponding highest excited state of the Hamiltonian $\Ham(T)$. In classic terms, the first case represents minimization of a cost function whereas the second involves its maximization.

Following \refeq{empty_memory_state}, the desired initial state for the considered heteronuclear two spin system is $\ket{\psi_{un}}=\frac{1}{2}(\ket{00} + \ket{01} + \ket{10} + \ket{11})$ with corresponding density matrix $\rho_{un} = \ketbra{\psi_{un}}$. Rewriting this density matrix in terms of spin operators, ignoring the identity term and proportionality constants, the initial state can be represented by the simplified and traceless operator $\rho(0) = \sigma^x_H + \sigma^x_C + 2 \sigma^x_H \sigma^x_C$. This state can be created starting from the thermal equilibrium using standard methods based on spatial averaging \cite{Knill1998}.

The transverse form of the initial Hamiltonian $\Ham_i$ suggests that pulse sequences applied on the transverse direction $x$ can be used not only to generate but also to effectively control the initial Hamiltonian. More specifically, $\Ham_i$ is represented by a transverse radio frequency (rf) field in positive or negative (depending on $w$) $x$ direction on both spins. The amplitude of the rf field $\Lambda(t)$ is reduced linearly from $\Lambda(0) = A_{max}$ to $\Lambda(T)=0$ during the fixed evolution time $T=50$ms which is shorter than relaxation times $T_1$ and $T_2$. The offsets $\nu_{H}$ and $\nu_{C}$ in \refeq{nmr_hamiltonian} were set to values $\pm 100$~Hz and $0$~Hz corresponding to values $\xi_i=\pm1$ and $\xi_i=0$, respectively.

In the simulations and experiments of quantum annealing, the time evolution of the system was discretized into $L$ steps of length $\Delta t = {T}/{L}$ with constant Hamiltonian $\Ham(t)$ for the time interval $[t,t+\Delta t]$. The continuous evolution is approximated for $L\rightarrow\infty$ and $\Delta t \rightarrow 0$. Our numerical results suggested that $L=100$ leads to sufficient accuracy. We chose $\Gamma=0.5$ for the weight parameter of the retrieval Hamiltonian term. We set $A_{max}=600$~Hz, where in simulations we get a reasonable value of $98\%$ for the overlap $\tr{\rho_f\rho_{exp}}$ between final state of the system $\rho_f$ and expected outcome state $\rho_{exp}$.

\begin{figure}
\includegraphics[width=8.5cm]{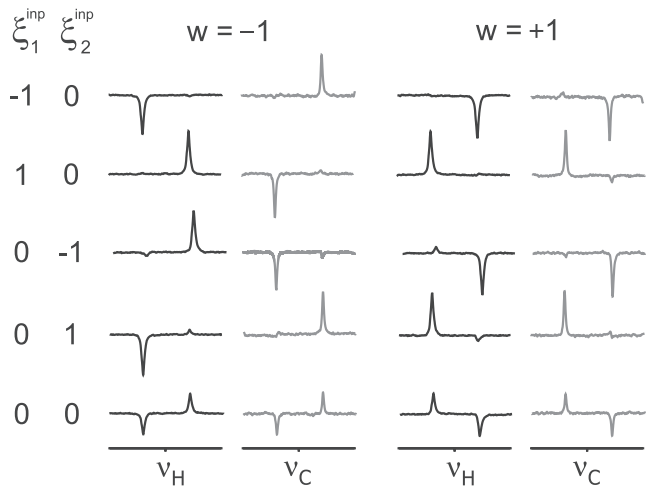}
\caption{Experimental $^1$H and $^{13}$C spectra for five combinations of $\xi^{inp}_{1}$ and $\xi^{inp}_{2}$ and two possible values of $w$, where $w=-1$ and $w=+1$ correspond to stored patterns with opposite and equal bit values, respectively. The spectral range in each spectrum is $\pm 240$~Hz.}
\label{fig:experiment}
\end{figure}

The result of the quantum annealing is read out by applying $\frac{\pi}{2}$ pulses followed by detection of the $^{1}$H or $^{13}$C signals. The $^{1}$H signals were acquired as single scan while $^{13}$C signals were acquired using 16 scans. The spectra recorded after the read out are shown in \reffig{experiment}. A reasonable match between the experimental results and simulations (data not shown) is found.
The first and third columns represent $^1$H spectra which encode the computational results as presented in \reftable{results}. The first four rows represent applications of associative pattern recognition with $n=1$ and $N=2$ where the algorithm evaluates the Hamming distance between the input and the partial memory patterns. A single peak at frequency $\nu_H\pm\frac{J}{2}$ corresponds to a unique recognized pattern. The peak's position and sign direction identify the pattern recognition results, e.g., left negative peak identifies the state $\ket{-1,1}$ corresponding to the output pattern [-1,1]. For the case of the blank input pattern $\xi^{inp}_{1}=\xi^{inp}_{2}=0$, the quantum neural network returns its memory content as consistent patterns [-1,1], [1,-1] for $w=-1$ and [-1,-1], [1,1] for $w=+1$. The resulting quantum state represents a superposition of pattern states identified by corresponding spectral peaks. For completeness, we show in the second and fourth column of \reffig{experiment} corresponding $^{13}$C spectra. Peaks at $\nu_C\pm\frac{J}{2}$ provide redundant information for the identification of the patterns.
\begin{table}
\begin{tabular}{c|c|c|c}
$\xi^{inp}_1$ & $\xi^{inp}_2$ & $\ket{\psi^{out}}$ for $w=-1$ &  $\ket{\psi^{out}}$ for $w=+1$\\
\hline
-1 & 0 & \ket{-1,\hspace{8pt}1} & \ket{-1,-1} \\
1 & 0 & \ket{\hspace{12pt}1,-1} & \ket{\hspace{11pt}1,\hspace{8pt}1} \\
0 & -1 & \ket{\hspace{12pt}1,-1} & \ket{-1,-1} \\
0 & 1 & \ket{-1,\hspace{8pt}1} & \ket{\hspace{11pt}1,\hspace{8pt}1} \\
0 & 0 & $\frac{1}{\sqrt{2}}\left(\ket{-1,1}+\ket{1,-1}\right)$ & $\frac{1}{\sqrt{2}}\left(\ket{-1,-1}+\ket{1,1}\right)$
\end{tabular}
\caption{The results of the pattern recognition represented by the corresponding pure quantum states.}
\label{table:results}
\end{table}

\section{Conclusion}

In summary, we have developed a new theoretical approach for quantum pattern recognition and implemented it using liquid state NMR techniques in a simple example with two qubits.
In the case of an associative memory, an incomplete input typically yields several memory patterns which are equally similar to this input. In contrast to classic neural networks, a quantum neural register can represent a superposition of recognized patterns. Whereas linearly combined classic mixture states, e.g., of the form $sgn(\pm \xi^\mu \pm \xi^\nu \pm \xi^\upsilon)$ \cite{Haykin1998} do not provide direct information about corresponding memory patterns, the quantum superposition states allow \textit{unambiguous} read-out of distinct patterns which form the superposition by means of an appropriate (iterative) measurement procedure.

The projector memory Hamiltonian enables the measurement of similarity between the input pattern and stored patterns as well as allows a differentiation between memory patterns and their reversed patterns $-\xi^\mu$ in contrast to classic Hopfield network (cf. \refapp{projector_memory}).
Since the problem is directly encoded into the final Hamiltonian $\Ham_p$, no ancilla qubits are required.
Furthermore, AQC dynamics rules out closed state cycles of the form $\S(t+T) = \S(t)$ with a period $T>1$ which are possible for classical neural networks \cite{Haykin1998}.

In the experiments, we realized the desired morphing between the Hamiltonians using quantum annealing. This allowed us to exploit the experimentally available control terms and did not require an iterative synthesis of the effective Hamiltonian trajectory \cite{Mitra2005}. Furthermore, we performed both ground and excited state AQC.

Several characteristics of the introduced concept remain open and require further research. The comparison of maximum storage capacity of the presented AQC based model and classic Hopfield network is an open problem that motivates additional study. Furthermore, time requirements for pattern retrieval can be investigated by studying the minimum energy gap during the course of the evolution \cite{Young2008} as well as employing heuristics related to success probability \cite{Farhi2001}.
Future investigations could also involve optimized annealing procedures which could provide significant improvements over alternative schemes \cite{Morita2008}.
An interesting enhancement of the proposed scheme could be introduced by considering asymmetric synaptic weights in neural modeling \cite{Chen2001}.
Future directions of research also include the extensions to larger spin systems using liquid state NMR quantum computation with the control of individual couplings of the effective Hamiltonian $\Ham(t)$ executed by means of pulse sequences and refocusing schemes \cite{Haeberlen1976, Khaneja2007, Sleator1995, DiVincenzo1998}.
The developed techniques can be potentially applied to alternative hardware designs, such as ion traps with magnetic field gradients \cite{Mintert2001} which have been already contextualized in quantum neural research \cite{Pons2007} and superconducting circuits \cite{Sporl2007}.

The authors thank Professor Wilhelm Zwerger at TU M{\"u}nchen for his support. R.N. and R.S. acknowledge support by Thomas Runkler and in particular Professor Bernd Sch{\"u}rmann who initiated the project at Siemens Corporate Technology. We thank Manoj Nimbalkar, Dr. Raimund Marx, and Dr. Wolfgang Eisenreich for their help in NMR experimentation. S.J.G. acknowledges support by the EU project QAP as well as by the DFG through SFB 631. R.N. acknowledges financial support from Siemens AG.

\appendix

\section{Projector memory Hamiltonian}

\label{sec:projector_memory}

In addition to the definition of the memory Hamiltonian of \refeq{memory_hamilt_coupl}, we introduce an alternative encoding of memory information based on a generalization of quantum search projection operators \cite{Roland2002}. It can be defined as a memory Hamiltonian
\begin{equation}
\label{eq:memory_hamilt_proj_a}
\Ham^{a}_{mem} = \Unit-\sum_\mu\ketbra{\xi^\mu}
\end{equation}
which is diagonal in the computational basis.
The definition represents the ideal case of memorized patterns encoded in the (degenerate) ground state energy of $\Ham^{a}_{mem}$ whereas not-memorized patterns correspond to excited states.
In combination with the input Hamiltonian $\Ham_{inp}$ \refq{retrieval_hamiltonian}, the bound for the input weight is given by $\Gamma < \frac{1}{2n}$ for arbitrary values of $p$ (cf. \refapp{input_weight}).

Alternatively, the memory Hamiltonian can be specified as a projector on the memory state
\begin{equation}
\label{eq:memory_hamilt_proj_b}
\Ham^{b}_{mem}=\Unit-\ketbra{\xi^{mem}}
\end{equation}
with memory state
\begin{equation}
\label{eq:stored_memory_state}
\ket{\xi^{mem}}=\frac{1}{\sqrt{p}}\sum_\mu\ket{\xi^\mu}.
\end{equation}
In the case of both projector memory Hamiltonians, the initial Hamiltonian can be chosen as a projector $\Ham_i = \Unit_N - \ketbra{\psi_{un}}$.

Using the definition $\Ham^{b}_{mem}$ \refq{memory_hamilt_proj_b}, we can solve the problem of quantifying the level of similarity between the input pattern $\xi^{inp}$ and all memory patterns. For this, the probability for measuring the memory pattern $\xi^\mu$ is interpreted as a measurement of relevance.
For comparison, the initial probabilities for the memory state \refq{stored_memory_state} are equally distributed with $p^\mu \equiv \sqnorm{\braket{\xi^\mu}{\xi^{mem}}} = \frac{1}{p}$.
In combination with the input Hamiltonian $\Ham_{inp}$ \refq{retrieval_hamiltonian}, the ground state of the problem Hamiltonian $\Ham_p = \Ham^{b}_{mem} + \Gamma \Ham_{inp}$ is
\begin{equation}
\label{eq:measure_similarity}
\ket{\psi^p_0} \propto \sum_\mu \frac{1}{\sqrt{p}} \left[ 1 - 2\Gamma \delta \hat{h}_\mu  \right] \ket{\xi^\mu} + O(\Gamma^2),
\end{equation}
where $\hat{h}_\mu=d_H(\xi^{inp},\hat{\xi}^\mu)$ is the Hamming distance between $\hat{\xi}^\mu=[ \xi^\mu_1,\ldots,\xi^\mu_n]$ and the retrieval pattern, $\delta \hat{h}_\mu = \hat{h}_\mu - \expect{\hat{h}}$ is the deviation of the Hamming distance from the average relative Hamming distance $\expect{\hat{h}}=\frac{1}{p}\sum_{\mu=1}^p \hat{h}_\mu$.
Thus, the input Hamiltonian shifts equally distributed weights in the memory state \refq{stored_memory_state} according to the corresponding Hamming distances.

We outline this effect using perturbation theory \cite{Messiah2000} for the problem Hamiltonian $\Ham^p$ in relation to the parameter $\Gamma$. The ground state energy and the corresponding eigenstate can be represented by
\begin{eqnarray}
\nonumber
E^p_0 &=& E^{mem}_0 + \Gamma E_0^{(1)} + \Gamma^2 E_0^{(2)} + O(\Gamma^3),\\
\label{eq:perturb_state}
\ket{E^p_0} &=& \ket{E^{mem}_0} + \Gamma \ket{E^{(1)}_0} + \Gamma^2 \ket{E^{(2)}_0} + O(\Gamma^3).
\end{eqnarray}
The first order approximation energy is given by $E_0^{(1)} = \braopket{E^{mem}_0}{\Ham^{inp}}{E^{mem}_0}$. Using the definition of the input Hamiltonian, it follows for $\hat{\xi}^\mu= [\xi^\mu_1,\ldots,\xi^\mu_n]$:
\begin{equation}
\label{eq:retriev_hamilt_on_stored_memstate}
\Ham^{inp}\ket{E^{mem}_0} = \frac{1}{\sqrt{p}} \sum_\mu \Ham^{inp} \ket{\xi^\mu} = \frac{1}{\sqrt{p}} \sum_\mu \left(-\hat{s}_\mu\right)\ket{\xi^\mu},
\end{equation}
where $\hat{s}_\mu = \xi^{inp} \cdot \hat{\xi}^\mu$ is the scalar product between vectors. Since $\hat{s}_\mu = n - 2\hat{h}_\mu$ with the Hamming distance $\hat{h}_\mu=d_H(\xi^{inp},\hat{\xi}^\mu)$ for the first $n$ elements, it follows
\begin{eqnarray*}
\Ham^{inp}\ket{E^{mem}_0} &=& \frac{1}{\sqrt{p}} \sum_\mu \left(-n+2\hat{h}_\mu\right) \ket{\xi^\mu}\\
&=& -n \ket{\xi^{mem}} + \frac{2}{\sqrt{p}} \sum_\mu \hat{h}_\mu \ket{\xi^\mu}.
\end{eqnarray*}
Using $\braket{\xi^\mu}{\xi^\nu} = \delta_{\nu\mu}$, we obtain for the first order energy correction
\begin{eqnarray*}
E_0^{(1)} &=& -n \braket{\xi^{mem}}{\xi^{mem}} + \frac{2}{p} \sum_{\nu\mu} \braopket{\xi^\nu}{\hat{h}_\mu}{\xi^\mu}\\
&=& -n + 2 \expect{\hat{h}}
\end{eqnarray*}
with average relative Hamming distance $\expect{\hat{h}}=\frac{1}{p}\sum_\mu \hat{h}_\mu$.

The first order approximation state is given by
\begin{equation*}
\ket{E^{(1)}_0} = -\sum_{k=1}^{2^N-1} \frac{\braopket{E^{mem}_k}{\Ham^{inp}}{E^{mem}_0}}{E^{mem}_k-E^{mem}_0} \ket{E^{mem}_k}.
\end{equation*}
By definition of the Hamiltonian $\Ham^{b}_{mem}$ \refq{memory_hamilt_proj_b}, the energy levels are $E^{mem}_0 = 0$ and $E^{mem}_{k>0} \equiv 1$. Since $\ket{E^{mem}_k}$ form an orthonormal basis,  it follows using \refeq{retriev_hamilt_on_stored_memstate}
\begin{eqnarray*}
\ket{E^{(1)}_0} &=& -\sum_{k\neq 0} \braopket{E^{mem}_k}{\Ham^{inp}}{E^{mem}_0} \ket{E^{mem}_k}\\
&=& -\left(\sum_{k\neq0}\ketbra{E^{mem}_k}\right)\Ham^{inp}\ket{E^{mem}_0}\\
&=& \left(\Unit - \ketbra{\xi^{mem}}\right) \frac{1}{\sqrt{p}} \sum_\mu \hat{s}_\mu \ket{\xi^\mu}\\
&=&  - \expect{\hat{s}} \ket{\xi^{mem}} + \frac{1}{\sqrt{p}} \sum_\mu \hat{s}_\mu \ket{\xi^\mu}
\end{eqnarray*}
with average relative scalar product $\expect{\hat{s}} = \frac{1}{p}\sum_\mu \hat{s}_\mu \equiv n - 2\expect{\hat{h}}$. Using state approximation \refq{perturb_state}, we obtain for the zero and first order
\begin{eqnarray*}
\ket{E^p_0} &=& \ket{E^{mem}_0} + \Gamma \ket{E^{(1)}_0}\\
&=& \frac{1}{\sqrt{p}} \sum_\mu \ket{\xi^\mu} + \Gamma \left(
\frac{1}{\sqrt{p}} \sum_\mu \hat{s}_\mu \ket{\xi^\mu} - \expect{\hat{s}} \frac{1}{\sqrt{p}} \sum_\mu \ket{\xi^\mu}
\right)\\
&=& \frac{1}{\sqrt{p}} \sum_\mu \left(1 + \Gamma\left(\hat{s}_\mu - \expect{\hat{s}}\right) \right) \ket{\xi^\mu}.
\end{eqnarray*}
We rewrite this expression in terms of the Hamming distance, thus obtaining expression \refq{measure_similarity},
\begin{equation*}
\ket{E^p_0} \propto \sum_\mu \frac{1}{\sqrt{p}} \left[ 1 - 2\Gamma \delta \hat{h}_\mu  \right] \ket{\xi^\mu} + O(\Gamma^2).
\end{equation*}
In case of all memory patterns $\xi^{\mu}$ being equidistant from the input pattern with regard to the first $n$ positions, the deviation $\delta \hat{h}_\mu$ equals zero so that corresponding measurement amplitudes are equally distributed. Otherwise, the amplitudes are shifted according to whether the distance to the input pattern is above- or below-average.

For completeness, we derive the second order approximation energy given by
\begin{equation*}
E_0^{(2)} = -\sum_{k=1}^{2^N-1} \frac{\left|\braopket{E^{mem}_k}{\Ham^{inp}}{E^{mem}_0}\right|^2}{E^{mem}_k-E^{mem}_0}.
\end{equation*}
It follows $E^{(2)}_0 = -4\Delta\hat{h}^2$ where $\Delta\hat{h}^2=\expect{\hat{h}^2}-\expect{\hat{h}}^2$ is the standard deviation for $\hat{h}_i$ and $\expect{\hat{h}^2}=\frac{1}{p}\sum_i \hat{h}_i^2$.\\

\begin{figure}
\includegraphics[width=8.5cm]{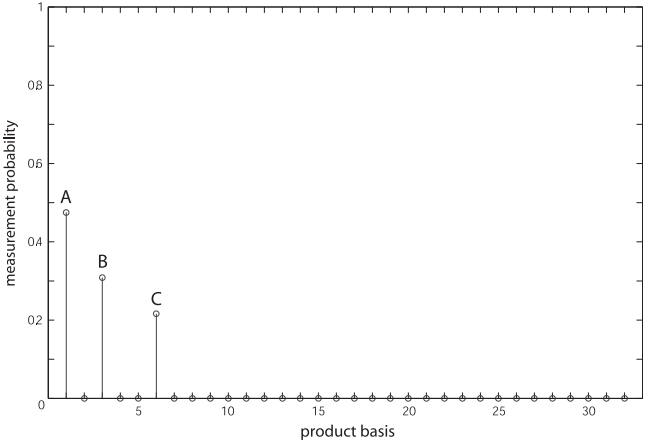}
\label{fig:similarity_measure}
\caption{Final measurement probabilities according to memory set \refq{similarity_measure_mem} and input pattern \refq{similarity_measure_inp}. The peaks correspond to memory patterns $\xi^1=\ket{-1,-1,-1,-1,-1}\equiv\ket{1}$ (peak A), $\xi^2=\ket{-1,-1,-1,+1,-1}\equiv\ket{3}$ (peak B) and $\xi^3=\ket{-1,-1,+1,-1,+1}\equiv\ket{6}$ (peak C) in binary and decimal representation, respectively.}
\end{figure}

In order to affirm the bias beyond the approximation to first order, we demonstrate the corresponding probability distribution for an exemplary memory set and input pattern
\begin{eqnarray}
\label{eq:similarity_measure_mem}
\P &=& \vect{rrrrrr}{
-1  & -1  & -1 & -1 & -1\\
-1  & -1  & -1 & +1 & -1\\
-1  & -1  & +1 & -1 & +1},\\
\label{eq:similarity_measure_inp}
\xi^{inp}&=&\vect{rrrrr}{-1  &  -1 &  -1  &  -1  &  -1},
\end{eqnarray}
with Hamming distances between the input pattern and the memory set given by $d_1 \equiv d_H(\xi^{inp},\xi^1) = 0$, $d_2=1$, and $d_3 = 2$.
\reffig{similarity_measure} shows for $\Gamma=0.1$ the distribution of measurement probabilities with $p_1\approx0.476$, $p_2\approx0.308$, and $p_3\approx0.216$.

\section{Input weight}

\label{sec:input_weight}

For $p=1$, we derive the bounds for the weight factor $\Gamma$ in combination with the coupling memory Hamiltonian $\Ham_{mem}$ \refq{memory_hamilt_coupl}. Since the Hamming distance between two patterns is symmetric under flips of the same positions, the memory pattern of length $N$ and input pattern of length $n \leq N$ can be written in the form
\begin{eqnarray*}
\xi^1 &=& [\overbrace {\underbrace{\: -1 \qquad  -1 \qquad \ldots \qquad -1}_{n}  \: \underbrace{-1 \quad \ldots \quad -1}_{N-n}}^{N}],\\
\xi^{inp} &=& [\overbrace{\underbrace{\; -1 \; -1 \; \ldots \; -1}_{n-m} \underbrace{+1 \; \ldots  \; +1}_m}^{n}],
\end{eqnarray*}
where $m$ is the Hamming distance between the patterns for the first $n$ elements.
An arbitrary vector $\xi$ of the length $N$ can thus be written in the form
\begin{eqnarray*}
\xi &=& [\overbrace{\underbrace{-1 \ldots -1}_{(n-m)-m_1} \underbrace{+1 \ldots +1}_{m_1}}^{n-m} \overbrace{\underbrace{-1 \ldots -1}_{m-m_2} \underbrace{+1 \ldots +1}_{m_2}}^{m}\quad\ldots\\
&&\qquad\qquad\ldots\quad\overbrace{\underbrace{-1 \ldots -1}_{(N-n)-m_3} \underbrace{+1 \ldots +1}_{m_3}}^{N-n}]
\end{eqnarray*}
where $m_1$, $m_2$, and $m_3$ represent the distances betweens $\xi$ and three respective segments of the memory pattern $\xi^1$ of the length $n-m$, $m$, and $N-n$.
Using the definition of memory and input Hamiltonians, we obtain
\begin{eqnarray*}
E^p(\xi^1) &=& E^{mem}(\xi^1) + \Gamma E^{inp}(\xi^1)\\
&=& -\frac{N}{2} + \frac{1}{2} - \Gamma\left(n-2m\right),\\
E^p(-\xi^1) &=& E^{mem}(-\xi^1) + \Gamma E^{inp}(-\xi^1)\\
&=& -\frac{N}{2} + \frac{1}{2} + \Gamma\left(n-2m\right),\\
E^p(\xi) &=& E^{mem}(\xi) + \Gamma E^{inp}(\xi)\\
&=& -\frac{1}{2N}\left[N-2M\right]^2 + \frac{1}{2}\\
\nonumber
&& -\Gamma\left(n-2\left(m_1+(m-m_2)\right)\right),
\end{eqnarray*}
where $M = m_1 + m_2 + m_3$ if $d(\xi,\xi^1) \leq d(\xi,-\xi^1)$ and $M = N-\left( m_1 + m_2 + m_3\right)$, otherwise.

Since $E^p(-\xi^1) - E^p(\xi^1) \propto \Gamma \left( n-2m \right)$, for all valid values of $\Gamma$ the memory pattern closer to the input pattern also has lower energy level. In the following, we assume $\xi^{inp}$ is closer to the original pattern $\xi^1$ with
$0 \leq m \leq \hat{r}$ where $\hat{r}=\lfloor \frac{n}{2} \rfloor$. For the system to return a valid answer, $\xi^1$ must be the global ground state of the problem Hamiltonian $\Ham_p$ \refq{problem_hamiltonian} with
\begin{equation}
E^p(\xi^1) < E^p(\xi\neq\xi^1).
\label{eq:hamilt_coupl_one_pattern_condition}
\end{equation}
From condition \refq{hamilt_coupl_one_pattern_condition}, it follows
\begin{equation}
\label{eq:tmp_condition}
\Gamma\left(m_2-m_1\right) < \frac{N^2-\left[N-2\left(m_1+m_2+m_3\right)\right]^2}{4N}.
\end{equation}
In order to obtain the upper bound for the value of $\Gamma$, we attempt to minimize the energy value corresponding to $\xi$. First, we assume $m_1=0$ so that the overlap between first segments of $\xi^1$ resp. $\xi^{inp}$ and the corresponding spins of $\xi$ is maximal. Furthermore, we assume $m_3 = 0$ so that the overlap between the third segment of $\xi^1$ and $\xi$ is maximal as well.

Since $m_2\leq m$ and $m$ is bounded by $\frac{n}{2}$, we obtain from \refeq{tmp_condition} the condition
\begin{equation*}
\Gamma < 1-\frac{n}{2N}.
\end{equation*}

A similar bound can be defined in combination with the projector memory Hamiltonian $\Ham^{a}_{mem}$ \refq{memory_hamilt_proj_a} for an arbitrary value of $p$. We consider a subset of the memory set $\P_{min}=\{\xi^1,\ldots,\xi^{p_{min}}\} \subseteq \P$, the patterns of which show the minimal distance to the input pattern $\xi^{inp}$ with $\forall \xi^\mu \in \P_{min}:\:d_H(\xi^{inp},\xi^\mu)=h_{min}$. The condition for the weight factor $\Gamma$ can thus be given with
\begin{equation*}
E^p(\xi^\mu) = -\Gamma\left(n-2 h_{min}\right) < E^p(\xi\neq \xi^\mu) = 1 - \Gamma \left(  n - 2 \hat{h} \right).
\end{equation*}
Since we are interested in a upper bound for $\Gamma$, we assume $\hat{h}=0$, i.e., $\xi$ completes the partial input vector. With $0 \leq h_{min} \leq n$, we obtain the corresponding condition
\begin{equation*}
\Gamma < \frac{1}{2n}.
\end{equation*}

\end{document}